\documentclass[11pt]{emulateapj}
\usepackage{graphicx}
\usepackage{multirow}
\usepackage{textcomp}
\usepackage{epsfig}
\usepackage{amsmath}
\usepackage{amssymb}
\usepackage{amsthm}
\usepackage{xy}

\def\3xmm{3XMM\,J1852$+$0033}

\newcommand{\swift}{{\em Swift}}

\def\ergs {erg\,s$^{-1}$}
\def\ergscm2 {erg\,s$^{-1}$cm$^{-2}$}

\def\cm2 {cm$^{-2}$}

\shortauthors{N. Rea et al.}

\begin{document}

\title{Constraining the GRB-magnetar model by means of the Galactic pulsar population}

\author{N. Rea\altaffilmark{1,2}, M. Gull\'on\altaffilmark{3}, J. A. Pons\altaffilmark{3}, R. Perna\altaffilmark{4}, M. G. Dainotti\altaffilmark{5}, J. A. Miralles\altaffilmark{3}, D. F. Torres\altaffilmark{2,6}}

\altaffiltext{1}{Anton Pannekoek Institute for Astronomy, University of Amsterdam, Postbus 94249, NL-1090 GE Amsterdam, the Netherlands.}         
\altaffiltext{2}{Instituto de Ciencias de l'Espacio (ICE, CSIC--IEEC), Campus UAB,  Carrer Can Magrans s/n, 08193 Barcelona, Spain.}
\altaffiltext{3}{Departament de Fisica Aplicada, Universitat d'Alacant, Ap. Correus 99, 03080 Alacant, Spain.} 
\altaffiltext{4}{Department of Physics and Astronomy, Stony Brook University, Stony Brook, NY, 11794, USA.} 
\altaffiltext{5}{Physics Department, Stanford University, Via Pueblo Mall 382, Stanford, CA, USA.} 
\altaffiltext{6}{Instituci\'o Catalana de Recerca i Estudis Avançats (ICREA), E-08010 Barcelona, Spain.}

\begin{abstract}

A large fraction of Gamma Ray Bursts (GRBs) displays an X-ray plateau phase within
$<10^{5}$\,s from the prompt emission, proposed to be powered by the spin-down energy of a rapidly
spinning newly born magnetar. In this work we use the properties of the Galactic neutron
star population to constrain the GRB-magnetar scenario. 
We re-analyze the X-ray plateaus of all \swift\, GRBs with known redshift,
between January 2005 and August 2014. From the derived initial magnetic field
distribution for the possible magnetars left behind by the GRBs, we study the evolution and properties of a simulated
GRB-magnetar population using numerical simulations of magnetic field evolution,
coupled with Monte Carlo simulations of Pulsar Population Synthesis in our Galaxy. We find that if the GRB X-ray plateaus are powered by the rotational energy of a newly formed magnetar, the current observational properties of the Galactic magnetar population are
not compatible with being formed within the GRB scenario (regardless of the GRB type or rate at z=0). Direct consequences would be that we should allow the existence of magnetars and "super-magnetars" having different progenitors, and that Type Ib/c SNe related to Long GRBs form systematically neutron stars with higher initial magnetic fields. We put an upper limit of $\leq$16 "super-magnetars" formed by a GRB in our Galaxy in the past Myr (at 99\% c.l.). This limit is somewhat smaller than what roughly expected from Long GRB rates, although the very large uncertainties do not allow us to draw strong conclusion in this respect.

\end{abstract}


\section{Introduction}

Gamma Ray Bursts (GRBs) are one of the most extreme and powerful
transient phenomena in the Universe. They are generally divided in two
groups, which have been proposed to have two distinctly different
origins: Long GRBs (LGRBs), connected to the Type Ib/c Core-Collapse
Supernovae, and Short GRBs (SGRBs), originating from the merger of
two neutron stars or a neutron star and a black hole.

Independently of the progenitor scenario, the prompt $\gamma$-ray
emission is followed by intense longer-wavelength emission
(afterglow). According to the standard "Fireball" theory, this
radiation arises from the formation of a blast wave, due to a
relativistic outflow pushing through the interstellar medium (Meszaros
\& Rees 1997; Sari, Piran \& Narayan 1998).  In the past decade,
thanks to \swift, the sample of Long and Short GRBs with a good
multi-band monitoring of the afterglow became sufficiently large to
enable a statistical study of the afterglow characteristics and
energetics (Nousek et al. 2006; O'Brien et al. 2006; Zhang et al. 2006; Willingale et al. 2007; Evans et al. 2009; Dainotti et al. 2011b, Margutti et
al. 2013, Dainotti et al. 2015b). It was observed that most GRBs do not show a smooth decay
in X-ray flux after the prompt emission, as expected for a pure
fireball model, but present rather ubiquitous X-ray plateaus at times
$<10^{5}$\,s, eventually pointing to a continuos energy injection in
the first hours/day after the GRBs. These X-ray plateaus are generally interpreted as due to: a newly-born rapidly-spinning magnetar (see i.e. Metzger et al. 2011), an accreting black hole (see i.e. Kumar et al. 2008) or a top-heavy jet evolution (Duell \& MacFadyen 2015). The similarity of these plateau phases between the two classes of GRBs was ascribed to a common
injection scenario. The fluence of these plateaus in both
LGRBs and SGRBs is comparable with that of the prompt emission (never
lower than an order of magnitude), and their luminosities and
durations are observed to be anti-correlated (Dainotti et al. 2008,
2010, 2011a,2013a; Rowlinson et al. 2013, 2014).

The latter correlation, combined with the fact that a newly born
magnetar could be formed either via the collapse of a massive star
(hence via a LGRB), or during the merger of two neutron stars (hence
via a SGRB), motivated the interpretation of these X-ray plateaus as
resulting from the delayed injection of rotational energy (with
$\dot{E}_{\rm rot}\sim10^{50}-10^{51}$ \ergs) from a fast spinning
magnetar \citep{usov92,zhang2001, metzger2011}.


\begin{figure*}
\centering
\vbox{
\includegraphics[width=8.3cm]{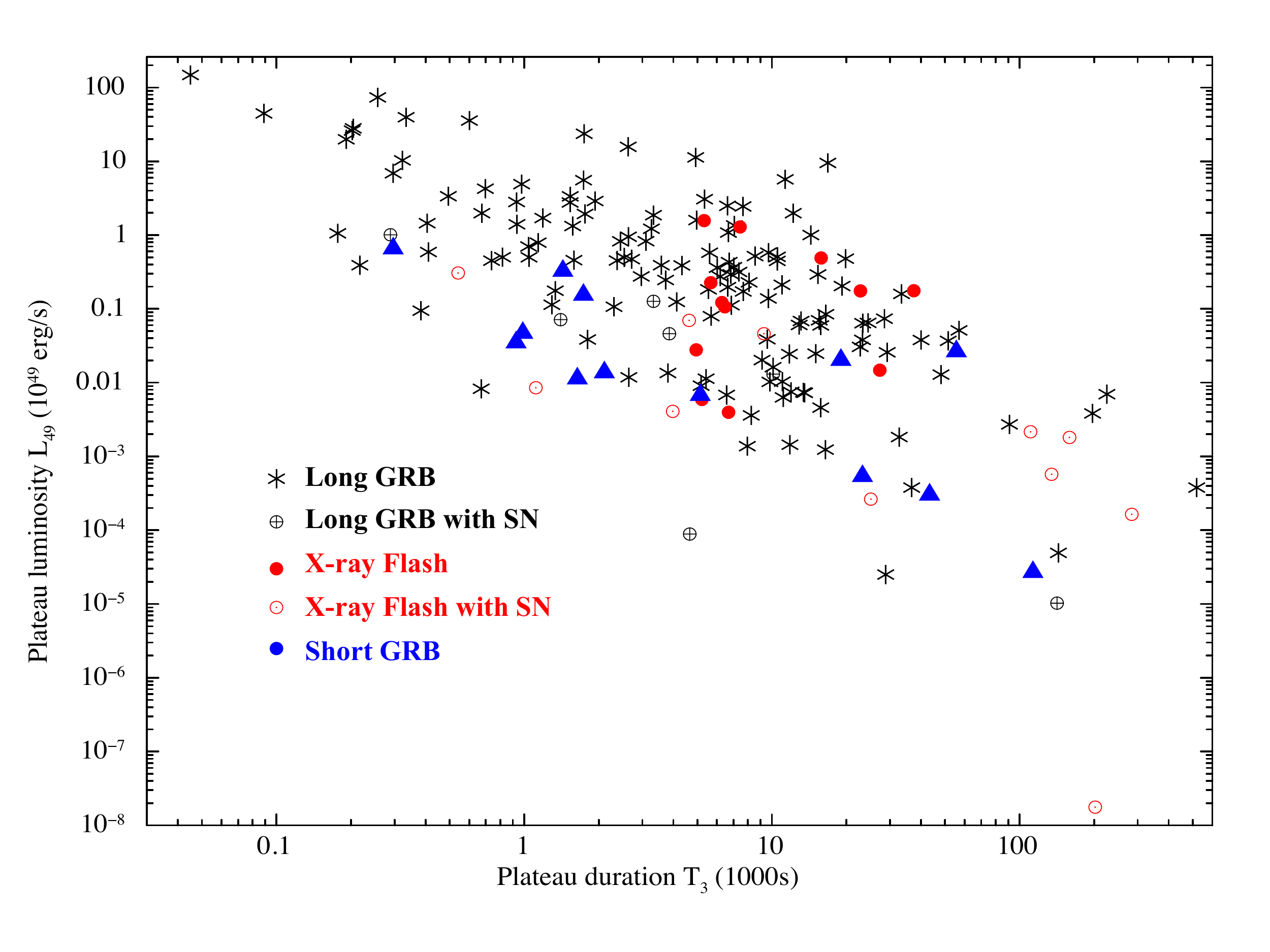}
\hbox{
\includegraphics[width=7.2cm]{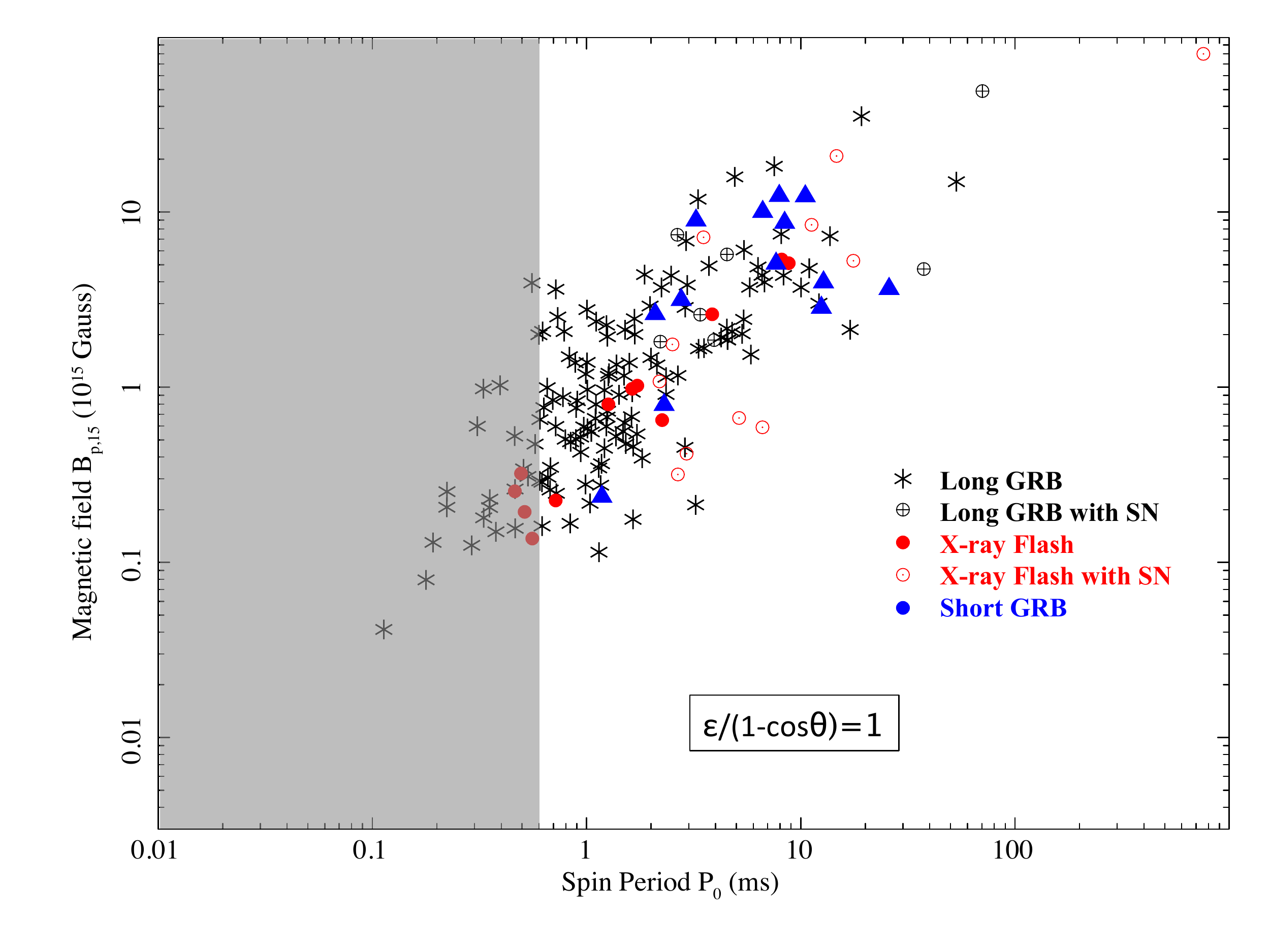}
\includegraphics[width=7.2cm]{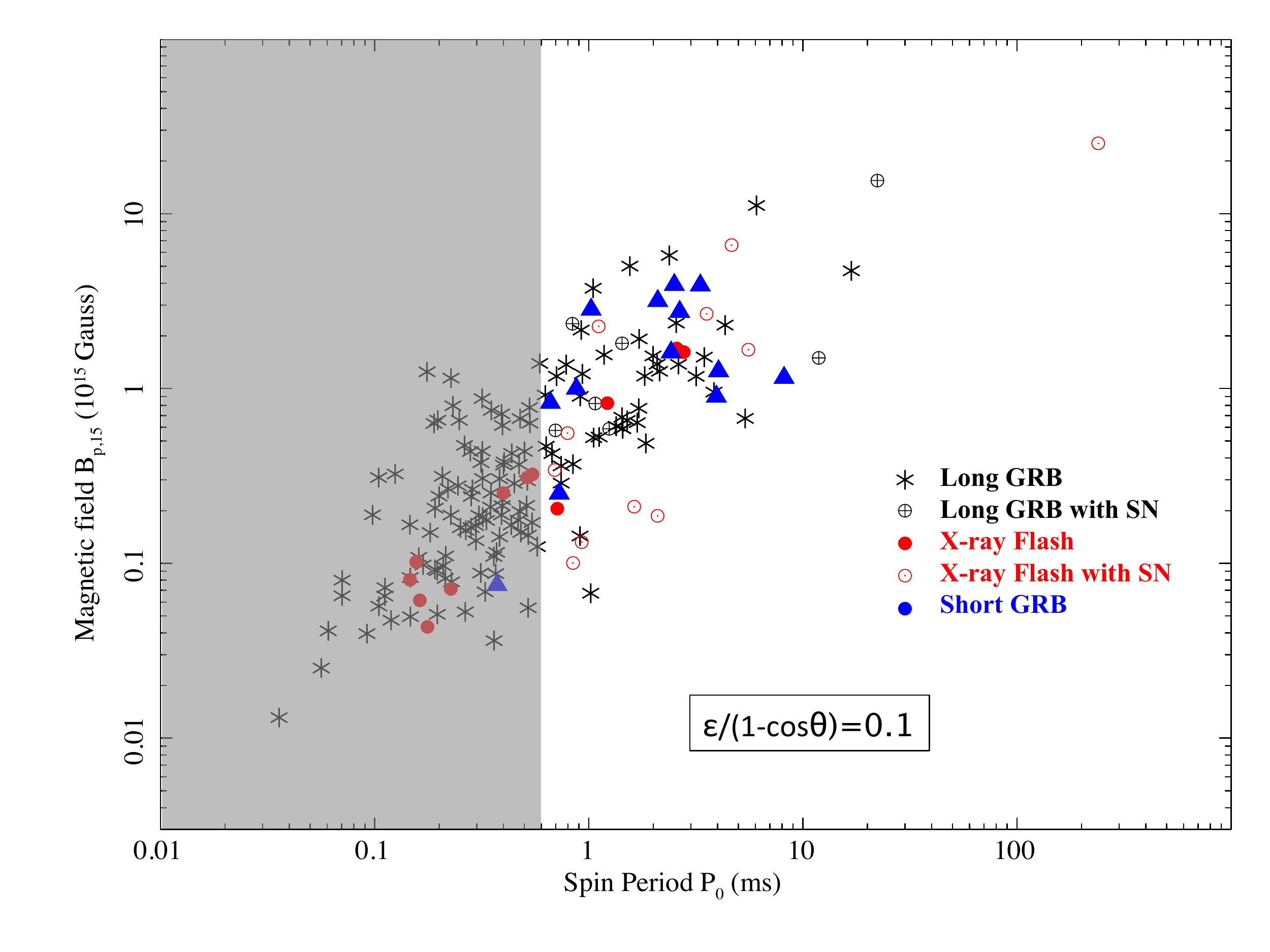}}}
\caption{{\em Top panel}: the rest-frame plateau durations versus the luminosity
  (1--10000 keV) at the end of the plateaus for all the GRBs in the
  sample (black = Long GRBs, Blue = Short GRBs, and Red = X-ray
  Flashes). {\em Bottom panels}: derived dipolar fields and initial spin
  period assuming the GRB-magnetar model for two different values of
  efficiency ($\epsilon$) versus opening angle ($\theta$) relation
  (see text for details). The shaded grey area excludes the rotational periods that would exceed the mass shedding limit under any reasonable neutron star EoS assumption.} \label{dataanalysis}
\end{figure*}


Further support to the GRB-magnetar scenario was provided by the
successful fitting of a large sample of Long and Short GRB afterglows
\citep{zhang2001, troja2007, lyons2009, dallosso11, bernardini2012,
  rowlinson2010, rowlinson2013, lu2014} by modeling the plateau duration and
luminosity in terms of the spin-down energy release timescale
($\tau_{\rm sd}$) and luminosity ($L_{\rm sd}$):
\begin{center}
$T_{\rm 3}\simeq\tau_{\rm sd}=2.05~(I_{45}B^{-2}_{p,15}P^2_{ms}R^{-6}_6) $\\
$L_{\rm 49}\simeq L_{\rm sd}=(B^2_{p,15}P^{-4}_{ms}R^6_6)$\;,
\end{center}
where $T_{\rm 3}$ is the plateau duration in $10^{3}$\,s, $L_{\rm 49}$
is the plateau luminosity in $10^{49}$ erg s$^{-1}$, $I_{45}$ is the
moment of inertia in units of $10^{45}$g~cm$^{2}$, $B_{p, 15}$ is the
magnetic field strength at the poles in units of $10^{15}$~G, $R_{6}$
is the radius of the neutron star in $10^{6}$\,cm, and $P_{ms}$ is the
initial period of the pulsar in milliseconds \citep{zhang2001,
  metzger2011}\footnote{These equations apply to the electromagnetic
  dominated spin down regime, since the gravitational wave dominated
  regime would be extremely rapid and produce a negligible
  electromagnetic signal. It is also assumed that the loss of
  rotational energy is given by the magneto-dipole formula, whose
  validity in this scenario is highly questionable.}. In this
scenario, the spin down luminosity and duration are expected to be
anti-correlated as: $\log(L_{\rm sd})=a - \log(\tau_{\rm sd}$), where
$a=\log(10^{52}~I^{-1}_{45} P^{-2}_{-3}$). Fitting the intrinsic
plateaus it has been obtained that $a=52.7\pm0.5$ and $L_{\rm 49}
\propto T^{(-1.07\pm0.14)}_{\rm 3}$ (Dainotti et al. 2013a; Rowlinson
et al. 2014). 

In our Galaxy we have discovered in the past few decades about 20
magnetars (Duncan \& Thompson 2002; see Mereghetti 2008, Rea \&
Esposito 2011 for recent reviews, and the McGill Magnetar
Catalog\footnote{http://www.physics.mcgill.ca/$\sim$pulsar/magnetar/}). They
are characterized by relatively bright X-ray luminosities ($L_{\rm
  X}\sim10^{33}-10^{35}$\ergs), rotational periods in the 0.3--12\,s
range, strong X/$\gamma$-ray flares and outburst activity, dipolar
magnetic fields in the $6\times10^{12}-10^{15}$\,G, and estimated ages
between $\sim 1-10^3$\,kyr.

In this paper we investigate the possible GRB origin of the magnetars
in our Galaxy, as well as derive the limits on the GRB-magnetar
scenario imposed by the properties of the Galactic magnetars. In
Sec.\,\ref{swiftanalysis} we re-analyze the {\em Swift} data of GRBs
with good redshift measurements. Fitting them with the GRB-magnetar
model (see also Rowlinson et al. 2014), we derive initial magnetic
fields and spin period distributions for the sample. In
Sec.\,\ref{popsyn} we use state-of-the-art magnetar evolution models
(Vigan\`o et al. 2013) coupled with Pulsar Population Synthesis
simulations (Gull\'on et al. 2014, 2015) to constrain the current
properties of possible magnetars formed via a GRB in our Galaxy in the
past Myr, by comparing synthetic populations with the observed
Galactic population. In Sec.\,\ref{discussion} we discuss our results as well as the issue of how
many and which GRBs are expected to leave behind a long-lived stable magnetar,
and the large uncertainties in the local GRB rates. We summarize our
results and draw conclusions in Sec.\,\ref{conclusion}.


\begin{figure*}
\centering
\vbox{
\includegraphics[width=8.4cm]{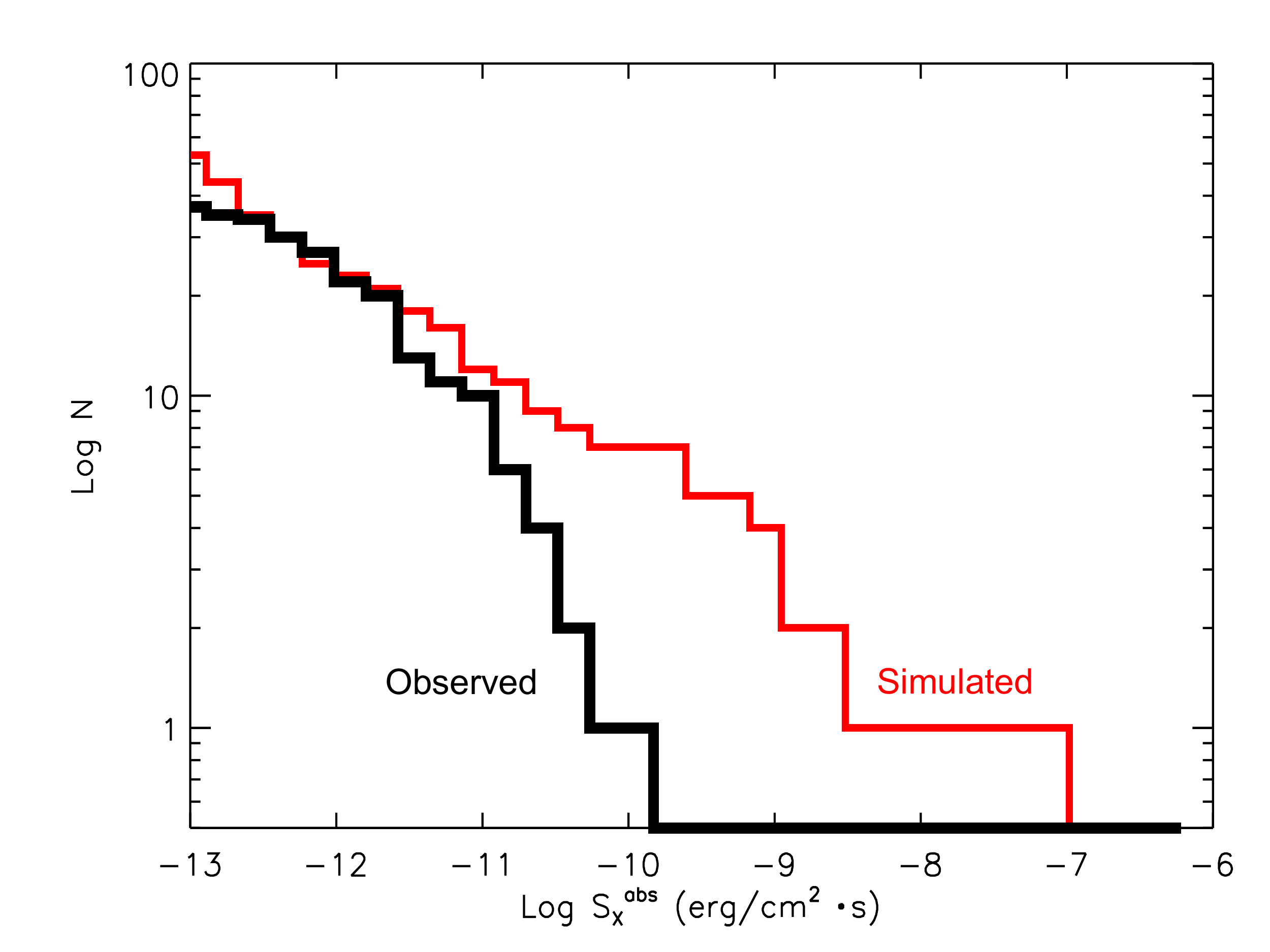}
\hspace{-0.47cm}
\hbox{
\includegraphics[width=8.0cm]{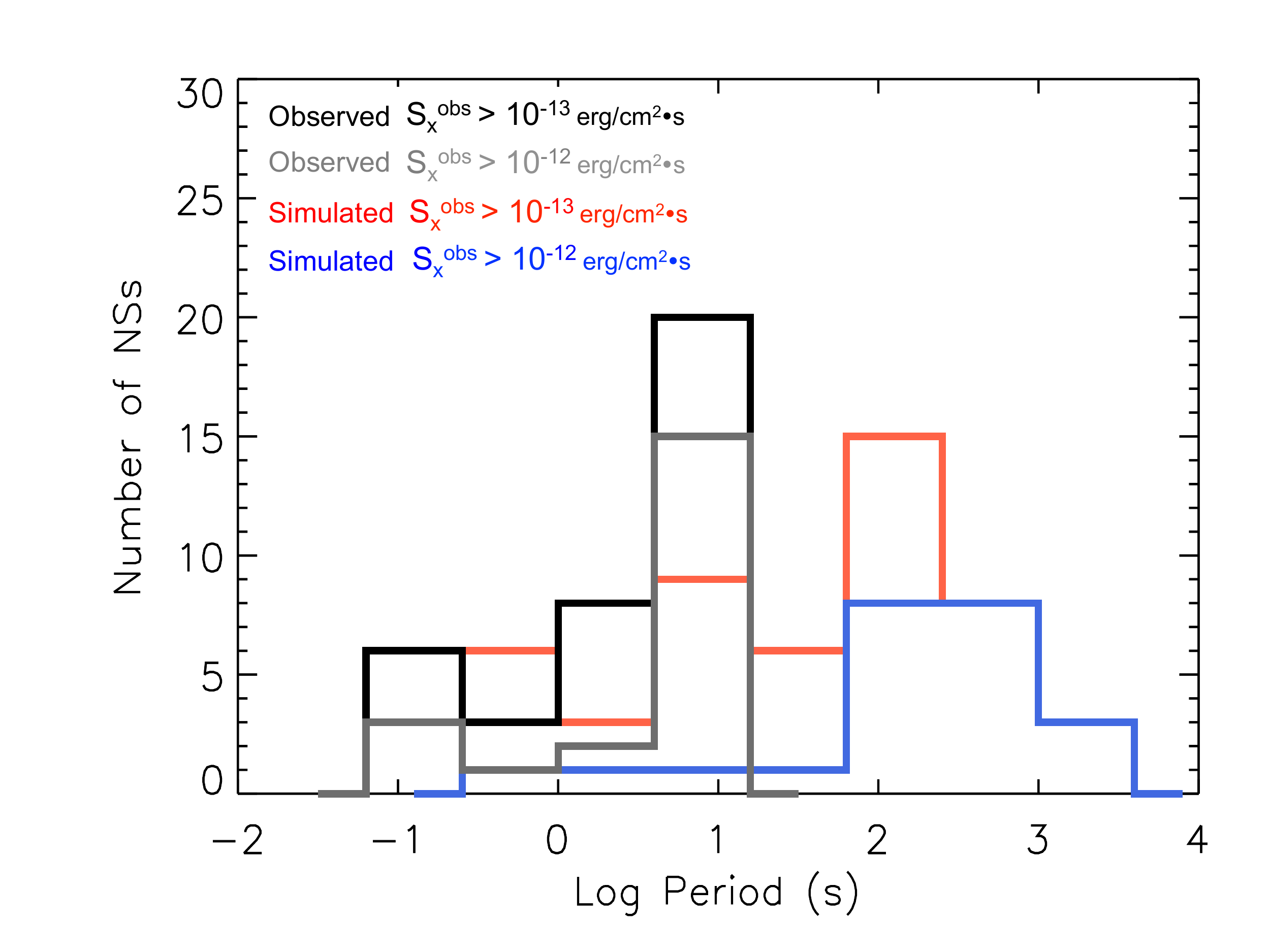}
\includegraphics[width=8.3cm]{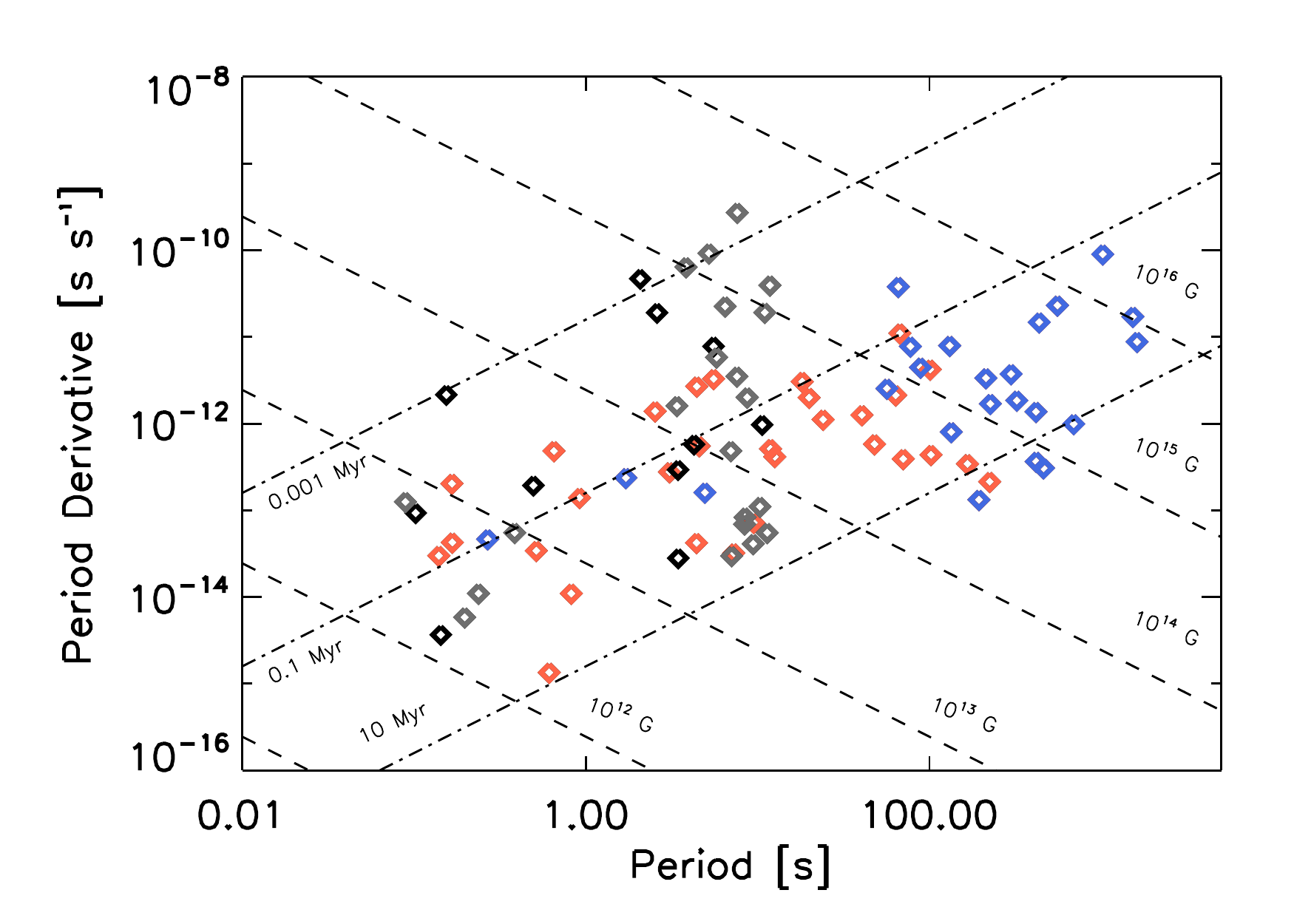}}}
\label{fig:ps}
\caption{Results of the Population Synthesis Simulations of our X-ray pulsar population plus 100
  "stable" magnetars formed via a GRB in the past Myr in our Galaxy. {\em
    Top panel}: logN--logS of the simulated sample (red) compared
  with the observed X-ray fluxes (black). {\em Bottom panels}: Spin period distribution of the observed X-ray pulsar population compared with the  "observable" simulated sample  of synthetic GRB-magnetars, and relative $P-\dot{P}$ diagram. In the latter, the black and grey symbols represent the observed objects, while the synthetic sample is displayed in red and blue. Grey and blue dots represent pulsars with X-ray fluxes of $>10^{-12}$\ergscm2 ; black and red dots are objects with fluxes  $10^{-13} < f_X < 10^{-12}$\ergscm2 .}

\end{figure*}

\section{Fitting magnetar-driven plateaus to the Swift light curves}
\label{swiftanalysis} 

We re-analyzed the sample of all GRB X-ray afterglows, detected by
{\it Swift} from January 2005 up to August 2014 with firm redshift
measurements, and for which the light curves include early X-ray data
and can be fitted by the Willingale et al. (2007) phenomenological
model. We followed the fitting procedure presented in Dainotti et
al. (2013a), and we use the redshifts available in the literature
\citep{X09}, in the Greiner web
page\footnote{http://www.mpe.mpg.de/~jcg/grbgen.html} and in the
Circulars Notice archive (GCN). The total number of GRBs with known
redshift (in the 0.033-9.4 range) observed by Swift until 2014 August
14th is $283$ (63 of which are SGRBs, and 25 are X-ray Flashes; XRFs),
but not all these GRBs show a well-defined X-ray plateau emission. We
found that among those, 176 GRBs afterglows (14 of which are SGRBs,
and all the 25 XRFs) are well fitted by an X-ray plateau model, as
described in Dainotti et al. (2013a). The fitting procedure fails
either when it gives unreasonable values of the errors or when the
determination of confidence interval in 1 $\sigma$ does not fulfill
the Avni (1976) prescriptions\footnote{http://heasarc.nasa.gov/xanadu/xspec/manual/}.

We plot in Fig.\,\ref{dataanalysis} (left panel) the rest-frame
plateau luminosities in the {\it Swift} bolometric band ($E_{min}$=1,
$E_{max}$=10000\,keV) at the time $T_a$ (end time of the plateau). The
luminosities are computed from: $L_X (E_{min},E_{max},T_a)= 4 \pi
D_L^2(z) \, F_X (E_{min},E_{max},T_a) \times \textit{K}$, where
$D_L(z)$ is the GRB luminosity distance (we have assumed a
$\Lambda$CDM flat cosmological model with $\Omega_M = 0.28$ and $H_0 =
70 {\rm km s}^{-1} {\rm Mpc}^{-1}$), $F_X$ is the measured X-ray
energy flux, $\textit{K}= (1+z)^{-1 +\beta_{a}}$ is the so called
\textit{K}-correction, where $\beta_{a}$ is the spectral index
assuming a simple power law spectrum (Evans et al. 2009; Dainotti et
al. 2010). Note that, in the current paper we use the variables
  $L_{49}$ and $T_{3}$ corrected for selection bias and redshift evolution (namely, de-evolved). This approach is slightly different from the one presented in Rowlinson et al. (2014): in this work we derive the slope of the correlation directly by using the de-evolved luminosity and time observables. In Rowlinson et al. (2014) the slope is fixed to the intrinsic one and the normalization is derived from the simulated data. This slightly different approach do not change the results, since the intrinsic slope used in Rowlinson et al. (2014) has been computed taking into account the same evolution. Caveats on the use
  of the observed slope instead of the intrinsic one have been discussed in
  Dainotti et al. (2013b). Since the rest frame time and luminosity we use are already corrected for selection bias and redshift evolution, the derived spin period and the magnetic field are unbiased too (this is a crucial point often omitted in the literature). In our analysis we have taken into account the undetected population of GRBs through the correction of the observed variables with the Efron \& Petrosian (1992) method.

In Fig.\,\ref{dataanalysis} (middle and left panels) we
then report on the inferred initial magnetic fields (at the neutron
star pole) and spin period distributions derived from modeling the
plateau luminosities and durations with a GRB-magnetar model (Zhang \&
Meszaros 2001), namely:
\begin{center}
$B^{2}_{0p,15}\simeq4.2025 I_{45}^{2}R^{-6}_{6}[L_{\rm sd, 49}*\epsilon/(1-\cos\theta)]^{-1}\tau_{\rm sd, 3}^{-2}$\\
$P^{2}_{0,-3}\simeq2.05 I_{45}[L_{\rm sd, 49}*\epsilon/(1-\cos\theta)]^{-1}\tau_{\rm sd, 3}^{-1}$ \, ,
\end{center}
where $\epsilon$ is the conversion efficiency of extracting rotational
energy from the newly born pulsar, and $\theta$ is the beaming
angle. We have studied the dependence of the derived initial $B_0$ and
$P_0$ distributions on these two unknown quantities. Lowering the
efficiency factor results in a net shift of the resulting B-fields and
periods towards lower values (with several GRBs requiring
unphysically low values of the birth rotational period; see
e.g. Fig.\,\ref{dataanalysis} right panel). On the other hand,
assuming an extremely beamed emission has the opposite effect,
shifting all inferred values toward longer spin periods but
unreasonably high magnetic fields (see also below). By studying the set of parameters that better reproduce the luminosity-time correlation, Rowlinson et al. (2014) propose a range for $\epsilon/(1-\cos\theta) \simeq 2-4$; this range leads to very high initial magnetic fields. For our purposes,
we adopt the less problematic case
$\epsilon/(1-\cos\theta) = 1$, but our conclusions will be unchanged if we assume larger values.

From the {\em Swift} X-ray plateau modeling we can derive a $B_0$
distribution for all GRBs, which we have distinguished in different classes. As it can be noted from
Fig.\,\ref{dataanalysis} (i.e. middle panel), there is no evidence for
a distinct $B_0$ and $P_0$ distributions as a function of the GRB
class. 


The $B_0$ field distribution of the resulting magnetars for all GRBs
is well fitted by a log-normal distribution centered at $\log{B_0} =
15.1$ with a dispersion of $\sigma \simeq 0.55$. Inferred rotational
periods at birth range between $\sim$ 0.1\,ms and 70\,ms (with a single outlier around 800\,ms). We note that
the fits give many unphysically short spin periods that would exceed the
mass shedding limit, which, depending on the neutron star mass, can be
placed between 0.6 and 1.5\,ms (see e.g. Goussard, Haensel \& Zdunik
1998).

\section{Neutron star population synthesis simulations and results}
\label{popsyn} 

In Gull\'on et al. (2015) we have performed a Population Synthesis
analysis considering both the radio-pulsar and the thermal X-ray
emitting neutron star populations (comprising the magnetars), taking
advantage of 2D numerical simulations of the magneto-thermal neutron
star evolution (Vigan\`o et al. 2013). We refer to Gull\'on et
al. (2014, 2015) for details on the Monte Carlo simulations used to
synthesize the Galactic neutron star populations. This analysis
allowed us to derive the best set of parameters ($B_0$ and $P_0$
distributions) consistent with both the current pulsar and magnetar
$P-\dot{P}$ distributions. The most important result of
this work was the discovery that a single log-normal B-field
distribution function could not explain at the same time the radio
pulsars and the magnetars. Either a truncated log-normal B
distribution, or a binormal distribution with two distinct
populations, were needed. More importantly, in both cases the current
lack of detected isolated X-ray pulsars with periods $>12$~s strongly
constrains the number of Galactic neutron stars born with
$B_0>10^{15}$~G .

We begin the simulations with the assumption of two different populations: normal
radio-pulsars, and magnetars associated
to GRBs. For the radio-pulsar population, we use the best fit
parameters corresponding to model D in Gull\'on et al. 2015, which
successfully fits the radio-pulsar population properties. The initial
magnetic field distribution for the synthetic magnetars is assumed to
be the one consistent with the GRB-magnetar association obtained in 
Sec.\,\ref{swiftanalysis} (a log-normal distribution centered at
$\log{B_0} = 15.1$ with dispersion $\sigma = 0.55$).  For every
synthetic magnetar, the initial period $P_0$ was forced to be
correlated with $B_0$ by the formula: $\log{P_0} = -6.2 + 0.22
\log{B_0}$, being the observed correlation between these quantities
intrinsic, and encompassing all different kind of GRBs (see
Fig.\,\ref{dataanalysis} and Dainotti et al. 2013a; Rowlinson et
al. 2014). We stress that the particular choice of $P_0$ is completely
irrelevant for our results, since the high average magnetic field of
the population makes them spin-down very fast to reach higher
periods. Assuming an initial period of 1 or 10 ms makes no difference
for the results discussed below.

The only parameter that still needs to be fixed is the relative normalization of the
number of radio pulsars and GRB-magnetars, the latter being expected to be proportional to the product of the GRB rate ($\rho_{\rm GRB}$) at
z=0, and the fraction of GRBs expected to leave as a remnant
compact object a "stable" magnetar ($f_{\rm mag}$; i.e. that survives subsequent collapse to a black hole). Given the large uncertainties in these two quantities, and the differences related to the different GRB types, we first run simulations and derive probabilities as a function of a general $\rho_{\rm  GRB}*f_{\rm mag} \equiv N_{gen}$: namely the number of "stable" magnetars that were formed via a GRB in the Milky Way in the past million years, regardless of the GRB type (this is allowed by the fact that all types have a similar $B_0$ distribution; see Fig.\,\ref{dataanalysis}). We then discuss differences in our conclusions for different GRB types in Sec.\,\ref{discussion}. Note that there is no GRB beaming
effect involved in the detectability of the remnant as an X-ray
pulsar, so there could be unseen GRBs leaving behind a visible
magnetar.  Initial positions in the Galaxy are drawn from the radial
probability distribution of Yusifov \& Kucuk (2004) within the
Galactic spiral arms. The position of each magnetar is then evolved
until its present age, by solving the Newtonian equations of motion
under the influence of the smooth Galactic gravitational potential
(Kuijken \& Gilmore 1989; Carlberg \& Innanen 1987). The age of each
star is randomly uniform in the interval $[0,1]$ Myr. Spatial kick
velocities and the inclination angle (between rotational and magnetic
axes) are also randomly selected. In order to account for the
detections in the X-ray band we assume blackbody emission plus
Resonant Compton Scattering, as typical of magnetars' spectra (Rea
et al. 2008; Zane et al. 2009). The photoelectric absorption along the
line of sight is also considered (see Gull\'on et al. 2015 for further
details).

In Fig.~\ref{fig:ps} we report the results of a typical realization
with $N_{gen} = 100$ magnetars (plus the large number of radio-pulsars
fitting the radio-pulsar population) by showing their predicted
$P-\dot{P}$ and $\log{N} - \log{S}$ diagrams for the visible X-ray
pulsars at present, compared with the observed sources. We show
results with two different cut-offs in absorbed X-ray fluxes, at
$10^{-13}$ and $10^{-12}$ erg s$^{-1}$ cm$^{-2}$. In the left panel,
we see that the total number of X-ray pulsars we observe in our Galaxy (after filtering for selection effects) is roughly consistent with the simulated radio pulsars plus 100
GRB-magnetars. This confirms our initial assumption of $\rho_{\rm  GRB}*f_{\rm mag} \equiv N_{gen} = 100$ in order to explain the currently observed $\sim$20 magnetars (after selection effects). However, it is clear that their distribution of X-ray
fluxes and spin periods are quite different from the observed
population. As expected, from the extremely high $B_0$ inferred from
the GRB plateaus, the simulated GRB magnetars are too bright and too
slow compared with the observed magnetars.  These discrepancies are
even more pronounced if we change the efficiency/beaming factor
($\epsilon/(1-\cos\theta)$) within the GRB-magnetar scenario, and
cannot be mitigated by changing the magneto-thermal evolutionary
model, or the crustal microphysics assumptions within reasonable
values (Vigan\`o et al. 2013; Pons, Vigan\`o \& Rea 2013).

Note that if some GRBs not showing a plateau phase still have magnetar central engines, i.e. with lower initial B-fields (hence with X-ray plateaus too faint to be detected over the afterglow), this will not change our conclusions, because the initial GRB-magnetar B-field distribution will not change systematically to lower values, but the log-normal distribution will only be slightly skewed to include also these putative additional GRB-magnetars with lower $B_0$.

Hence, our first result is that our observed population of magnetars
cannot be formed via a GRB (regardless of the assumptions on the rates or the different GRB types) because they would have
luminosity and period distributions largely inconsistent with the
observational data.

We can now estimate the maximum number of "stable" GRB-magnetars in the Milky Way left behind in the past
Myr, that is still compatible with the observations.

As shown in Gull\'on et al. 2015, the number of detectable synthetic
magnetars in each realization closely follows a Poissonian
distribution with mean value $\lambda = N_{gen} * p$, where $p$ is the
detection probability (note that we assume the GRB rate constant over 1\,Myr timescale). We have calculated this probability by
performing a large number of runs ($\sim 500$ realizations), and we
obtained $p=0.18$ and $p=0.28$ for fluxes $>10^{-12}$ and $>10^{-13}$,
respectively.

\begin{figure}
\centering
\includegraphics[width=9cm]{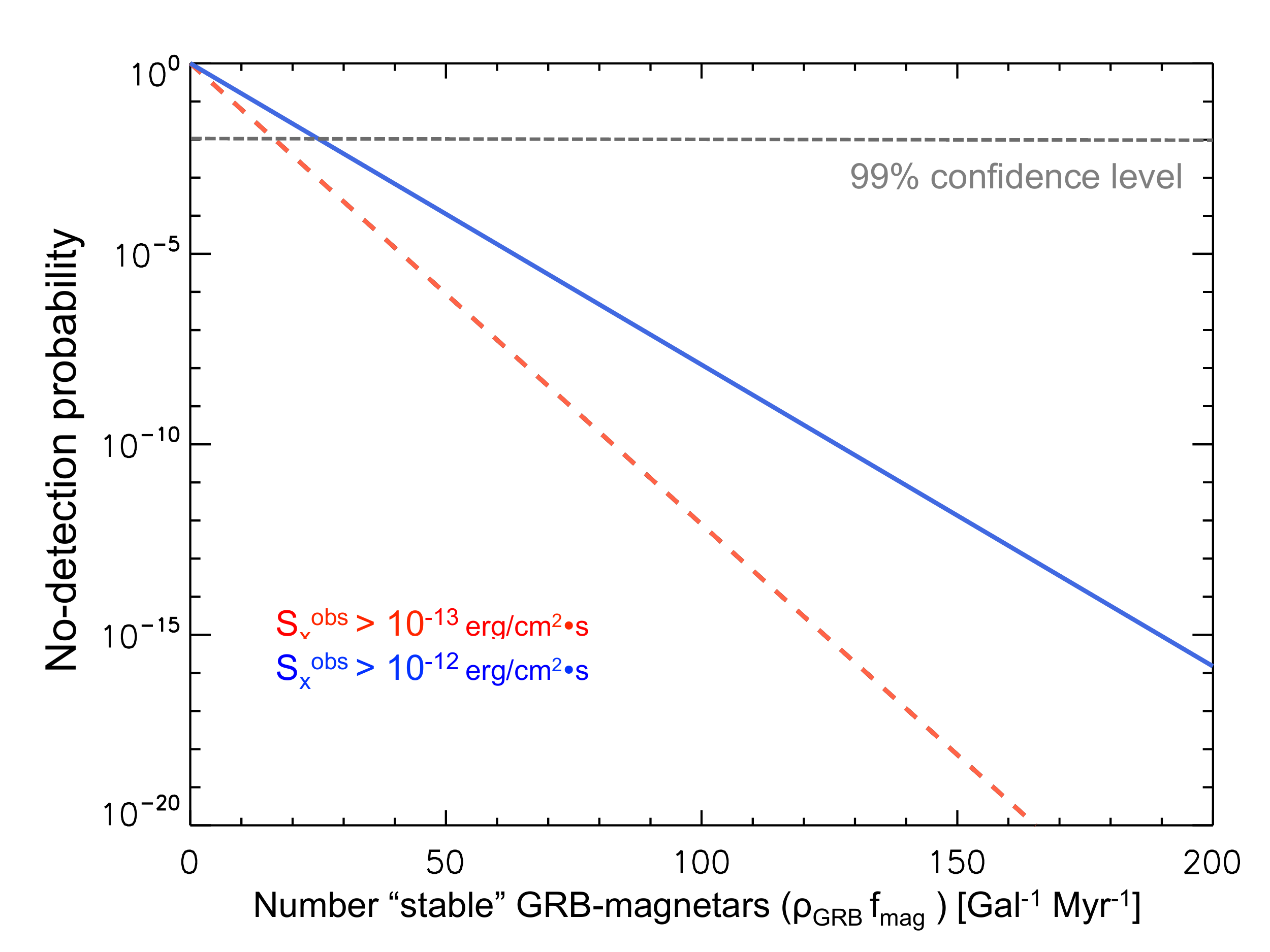}
\label{prob}
\caption{Probability of the no-detection of a synthetic magnetar with a spin period $>12$\,s as a function of the injected number of "stable" magnetars for two cuts fluxes: $10^{-13}$ (dashed red) and $10^{-12}$ (solid blue) erg s$^{-1}$ cm$^{-2}$. The grey line indicates the 99\% confidence level.}
\end{figure}

Since the most constraining observational fact is the lack of X-ray
pulsars with periods greater than 12\,s, we can calculate the probability of not detecting any pulsar with $P>12$\,s, which is $e^{-\lambda }$.  In
Fig.~\ref{prob} we plot the no-detection probability of magnetars with
spin period $>12$\,s as a function of $N_{gen}$, the number of
"stable" magnetars formed in the Milky Way via a GRB in the past Myr. The figure compares
the results for two different flux thresholds, $10^{-13}$ erg s$^{-1}$ cm$^{-2}$ (dashed red) and
$10^{-12}$ (solid blue) erg s$^{-1}$ cm$^{-2}$. These two
fluxes roughly encompass the range in which we believe our X-ray sample
of detected X-ray pulsar is nearly complete. Thus, assuming our sample
is complete above fluxes $>10^{-13}$ erg s$^{-1}$ cm$^{-2}$, only in
one per cent of the simulations we do not find any visible pulsar with
$P>12$\,s, which means that we can establish the upper limit $\rho_{\rm
  GRB} * f_{\rm mag} < 16$ with a 99\% confidence level. 
  
The above conclusions are in principle valid for any GRB type leaving behind a "stable" magnetar. However, as we will discuss further in the following section, the upper limit we derived is eventually meaningful only for LGRBs, given that SGRBs are hardly expected to leave any "stable" magnetar, and expected to collapse into a black hole soon after the X-ray plateau phase in most of the cases.


\section{Discussion}
\label{discussion}

We have performed neutron star population synthesis simulations, to set
constraints on the GRB-magnetar scenario by means of the Galactic
population of highly magnetized neutron stars. By assuming that the
X-ray plateau phases of GRB afterglows are powered by the rotational
energy of a newly born, rapidly spinning magnetar central engine, we
derived from the {\em Swift} GRB sample the resulting initial B-field
and spin period distribution of such newly born magnetars. Using these
distributions, we simulated the properties of a synthetic population
of magnetars formed in our Galaxy in the past Myr via a GRB.

We found that, if we assume  $\sim$100 GRBs
leaving behind a "stable" magnetar in the past Myr, the number of
"observable" objects (considering the predicted properties of such
simulated magnetars and all the observational biases) roughly agrees
with the number of magnetars we currently observe in the Milky Way
($\sim20$). However, the properties of the simulated sample of
GRB-magnetars are inconsistent with what observed: they are 
too bright, and spin too slowly, with spin periods far exceeding the
observed limit of 12\,s for the Galactic pulsar population (see Pons,
Vigan\`o \& Rea 2013).

This result was not totally unexpected given that, to model the
current population of pulsars and magnetars in our Galaxy, it was
recently observed that the initial $B$-field distribution 
should not allow fields in excess of $10^{15}$\,G,
otherwise the limiting spin period observed in isolated pulsars
($\sim12$\,s) cannot be reconciled (Gull\'on et al. 2015; Popov
2015). The magnetars that GRBs need to form to supply spin-down energy to the X-ray plateaus, are "super-magnetars", having initial B-fields significantly larger than those extrapolated for our Galactic magnetars.

Given that the number of observable magnetars is reproduced, but their
general properties are not, we can safely conclude that assuming the
GRB-magnetar scenario in its present formulation, in particular that X-ray plateaus are powered by spin-down energy, our Galactic
magnetars (regardless the assumed GRB type or rate at z=0) should be mostly formed by a distinct formation path than a GRB. Most likely a type of Core-Collapse SNe different from the Type Ib/c connected to Long GRBs. In this contest, this would also mean that GRB-like SNe should systematically produced stronger magnetic fields in the proto-magnetar than other CC-SNe\footnote{Unfortunately assessing whether this is or not the case is currently beyond the capabilities of current simulations of magnetic field formation in proto-neutron stars (and certainly far from the aim of this work).}.

\subsection{General estimates of the fraction of expected stable magnetars ($f_{\rm mag}$), and GRB rates at z=0}
\label{rates}

To put our simulations in contest, we discuss here current estimates of the local GRB rates, and of the probability for a magnetar born associated to a
GRB to survive or collapse to a black hole after the X-ray
plateau phase. Note that both these quantities are extremely uncertain.

If we proceed observationally to derive $f_{\rm mag}$, within the GRB-magnetar model, we can assume that if an X-ray plateau is observed the GRB formed a magnetar. 
From the {\em Swift} GRB reanalysis we derived that, for LGRBs, in 70\%
of the cases we can reasonably fit a plateau phase (137 cases over
195), for SGRBs, a plateau improves the fit in 23\% of the cases, and for
XRFs in 100\%.  We then assume this percentage as the minimum percentage of
GRBs having a magnetar engine powering the plateaus (in the others the plateau could had been missed or too faint).

Subsequently, we assume to zeroth order that if there is no collapse onto a black hole
(i.e. due to residual accretion onto the newly formed magnetar), at
the end of the X-ray plateau there is no sharp decay in time, and the
afterglow decays as $t^{-\alpha}$, with $\alpha\leq2$.
To estimate the fraction of magnetars that collapse, we have
counted in how many cases we found a subsequent
$t^{-\alpha}$ decay with $\alpha > 2$. We find that such steep
decay after the X-ray plateau is detected in 14 LGRBs, among the 137 with an X-ray plateau. We can then roughly estimate
that the fraction of LGRBs leaving a "stable" magnetar is $f_{\rm mag}\sim0.63$.

Regarding the determination of the LGRB rates at z=0, two main approaches
have been discussed in the literature (in addition to the limits
inferred via radio afterglow constraints; \citealt{perna98}). The first approach derives
the local LGRB rate from the GRB association with SN Type Ib/c. Radio and
optical SN surveys suggest that $\sim25$\% of all CC-supernovae are
Type Ib/c, but only 3-10\% of those are related to LGRBs (Berger et
al. 2003; Soderberg et al. 2006, 2010; Li et al. 2011; Lien et
al. 2014). In the local Universe this type of SNe have a rate of
$\rho_{\rm SN-Ib/c} = 1.7\times10^{4}$\,Gpc$^{-3}$ yr$^{-1}$
(Cappellaro et al. 1999, Soderberg et al. 2010). With one galaxy in
100 Mpc$^3$ (or equivalently, with the Milky Way volume of about
$10^{-7}$\, Gpc$^3$; Panter et al. 2007) this results 
in $\sim$50--170 GRB-SN Ib/c events within the last million years.

A different approach relies on a direct inversion of the
redshift-luminosity distribution of the observed LGRBs to infer their
local rate. This method needs to impose a low-luminosity cutoff to
avoid divergences (see e.g. discussion in
\citealt{guetta04}). A comprehensive study with {\em Swift} GRBs up to 2010
was performed by Wanderman \& Piran (2010). With a low-$L$ cutoff of 
$L > L_{50}\equiv10^{50}$\ergs, they inferred a local rate of  $\rho_{{\rm LGRB}(L>L_{50})} = (1.3\pm0.6) f^{-1}_b$
(Gpc$^{-3}$ yr$^{-1}$). 
Correcting for a beaming factor of about $f^{-1}_b = 70$ (see Guetta
et al. 2005; Fong et al. 2012), we expect $\sim$9 LGRB$_{L>50}$ events
in 1 Myr. We caveat here that there might be a
  metallicity dependence in extrapolating this GRB rate at z=0 (in particular Milky Way-like galaxies seem not to be the preferred hosts for LGRBs; see e.g. \citealt{robertson12, salvaterra12,trenti13,trenti15}). 
However, while some evidence points towards a preference of LGRBs for low-metallicity hosts 
(e.g. \citealt{modjaz08,graham13}), some outliers
  have also been discovered (Savaglio et al. 2012; Levesque 2014). 
The uncertain dependence of the GRB rate on metallicity and star formation, as well as on redshift, only
contributes to increase the uncertainties of the local LGRB rate determination (Jimenez \& Piran 2013; Dainotti et al. 2015). For the above reasons we do not enter in the metallicity/redshift/star formation rate dependence discussion, especially because it is not so relevant for the work presented here.

As Wanderman
\& Piran (2010) discuss in their Sec. 6.2, there are several
low-luminosity LGRBs$_{L<50}$ that are
not taken into account in their estimated rate. Given their faint nature, LGRBs with $L<L_{50}$
could have a rate much larger than for brighter LGRBs,
but at this time it remains even more uncertain. Current estimates state that they should be
roughly 10 times more numerous than the LGRBs$_{L>50}$ (Soderberg et
al. 2006a, 2010), and have very low beaming factors. Guetta \&
Della Valle (2007) attempted to estimate their local rates
on the basis of the few known events, and inferred
$\sim380_{-225}^{+620}$(Gpc$^{-3}$ yr$^{-1}$), which would result in
about 38 low-luminosity LGRB events in the past Myr (again with large
errors). This is consistent with a similar estimate found by Liang et al. (2007), assessing the rate of the low-luminosity LGRBs as $\sim0.7$\% of all Type Ib/c SN.

Summarizing, the different approaches estimate that the total (very uncertain), beaming corrected, LGRBs rate at z=0 should range within $\sim$50--170 (considering also the low luminosity ones) in our Galaxy in the past Myr, depending on the different approaches in the literature.

For SGRBs, Wanderman \& Piran (2015) derived $\rho_{\rm SGRB} = (4\pm2
)f^{-1}_b$(Gpc$^{-3}$ yr$^{-1}$), where $f^{-1}_b$ is the GRB beaming
factor. Assuming $f^{-1}_b = 30$ (see Fong et al. 2012), we then
expect $\sim$12 SGRB events in our Galaxy in the past \,Myr.  The
estimate of $f_{\rm mag}$ for SGRBs is even more difficult than for
LGRBs. Observationally, this is very much limited by the smaller sample to be meaningful. On the other hand, 
theoretically, while it has been demonstrated via general
relativistic, magnetohydrodynamical simulations that the formation of
a stable neutron star from the merger of two small neutron stars
($\sim1.2$M$_{\odot}$) is possible \citep{giacomazzo13,dallosso15}, the formation
rate depends on the rate at which the small-mass neutron stars are
formed at birth, as well as on the neutron star equation of state
(which determines the maximum mass of the resulting magnetar), and on the magnitude of the subsequent rate of accretion.

All in all, since the statistics of neutron star masses in binaries are
still too small to draw quantitative estimates, and the local,
galactic SGRB rates are smaller than those of LGRBs anyways, we have
adopted the conservative assumption that the possible contribution
from SGRBs to the observed galactic magnetar population is negligible (note also that Galactic magnetars are mostly located in the Galactic plane and in massive star clusters, unlike what would be expected for the remnants of a compact merger). Hence, even if our results are not dependent on the GRB type, but require only such GRB to leave a "stable" magnetar behind, eventually our conclusions and constraints are meaningful only for the LGRB population.

\subsection{Constraints on Long GRBs}

With our simulations we have also estimated the probability of
non-detecting a GRB-formed magnetar in our current population as a
function of the number of "stable" magnetars that a GRB, mainly LGRBs, have left in the galaxy in the last
Myr, namely  $\rho_{GRB} * f_{\rm mag}$.  We find that, in order to reconcile at a 99\% confidence 
level the non-detection of a GRB-magnetar compact remnant in our Galaxy (meaning non-detecting any magnetar with $P>12$\,s), the quantity $\rho_{GRB} * f_{\rm mag}$ should not
exceed $\leq16$ Gal$^{-1}$\,Myr$^{-1}$. This number depends
mainly on the completeness of the X-ray sample of observed neutron
stars, hence it can be revised further, and become more stringent, with the advent of new deep
X-ray surveys such as eROSITA (Merloni et al. 2012).

Extrapolating current LGRB rate estimates, we derived rough values of
$\rho_{LGRB}\sim50-170$\,Gal$^{-1}$\,Myr$^{-1}$, and $f_{\rm
  mag}\sim0.63$ (from fitting the \swift\, data), that result in
$\rho_{LGRB} * f_{\rm mag} \sim 30-110$. This is somewhat larger (although with large uncertainties) than the maximum allowed number of GRB "super-magnetars" in our Galaxy ($< 16$ at 99\% confidence level).

\section{Conclusions}
\label{conclusion}

Our results show that the initial B-field distribution needed to explain the GRB X-ray plateaus in terms of a fast spinning magnetar does not reconcile the properties of these GRB-magnetars with our Galactic magnetar population, even using the most favorable choices of efficiency/beaming factors. We should then allow the existence of magnetars and "super-magnetars", with two different progenitors and formation path, and different magnetic field formation efficiency.  

Even though the large uncertainties in the GRB rates at z=0, in the metallicity and star formation rate dependences, and in the fraction of neutron stars collapsing to a black hole, do not allow anyhow to rule out the GRB-magnetar model on the basis of the observed Galactic population, several fine-tunings are needed to maintain the model in its present form, and keeping the interpretation that X-ray plateaus are necessarily due to spin-down energy (i.e. we should allow some progenitors or environments to create systematically more magnetic stellar remnants than others). 

If those stable GRB-formed "super-magnetars" indeed exist, their current non-detection in our Galaxy can be used to put limits on $\rho_{LGRB} * f_{\rm mag}$, that will get possibly more and more constraining by means of future deep X-ray surveys.

\vspace{0.5cm}

\acknowledgements
NR thanks A. Rowlinson, B. Metzger, R. Margutti, B. Zhang, S. Dall'Osso, A. Soderberg, R. Wijers, A. MacFayden, M. Modjaz, S. Campana, P. D' Avanzo, M.~G. Bernardini, and L. Rezzolla for useful discussions and/or comments on the manuscript, and the referee for the careful reading. NR is supported by an NWO Vidi Grant, and kindly acknowledges Harvard ITC, NYU, Stony Brook University and the MIAPP institute in Garching, for the hospitality during the preparation of this work. NR and DFT are also supported by grants
AYA2012-39303 and SGR2014-1073. MG is supported by the fellowship
BES-2011-049123. MG, JAP and JAM acknowledge support by grants AYA2013-
42184-P and Prometeu/2014/69. RP acknowledges support from NSF grant No. AST 1009396. M.G.D. is supported by FP7-PEOPLE-2013-IOF under the grant agreement number 626267, and thanks the ITHES group and the Astrophysical Big Bang Laboratory for fruitful discussions.  This work is partially supported by the European COST Action MP1304 (NewCOMPSTAR).

\label{lastpage}


\begin{thebibliography}{30}



\bibitem[Avni(1976)]{1976ApJ...210..642A} Avni, Y.\ 1976, \apj, 210, 642 

\bibitem[\protect\citeauthoryear{Berger et al.}{2003}]{berger2003} Berger E., et al., 2003, Nature, 426, 154 

\bibitem[\protect\citeauthoryear{Bernardini et  al.}{2012}]{bernardini2012} Bernardini M.~G., Margutti R., Mao J., Zaninoni E., Chincarini G., 2012, A\&A, 539, A3 

\bibitem[Cappellaro et  al.(1999)]{1999A&A...351..459C} Cappellaro, E., Evans, R., \& Turatto, M.\ 1999, \aap, 351, 459 

\bibitem[Carlberg \& Innanen(1987)]{1987AJ.....94..666C} Carlberg, R.~G., \& Innanen, K.~A.\ 1987, \aj, 94, 666 

\bibitem[\protect\citeauthoryear{Dainotti et al.}{2008}]{dainotti2008} {Dainotti}, M.~G., {Cardone}, V.~F., {Capozziello}, S., 2008, MNRAS, 391, 79

\bibitem[\protect\citeauthoryear{Dainotti et al.}{2010}]{dainotti2010} Dainotti, M.~G., Willingale, R., Capozziello,  S., Cardone, V. F., Ostrowski M., 2010, ApJ, 722, L215 

\bibitem[\protect\citeauthoryear{Dainotti et al.}{2011a}]{dainotti2011a} Dainotti, M.~G., S., Cardone, V. F., Capozziello, S. Willingale, R., \& Ostrowski M., 2011, ApJ, 730, 135 

\bibitem[\protect\citeauthoryear{Dainotti et al.}{2011b}]{dainotti2011b} Dainotti, M. G., Ostrowski, M. \& Willingale, R., 2011, MNRAS, 418,2202

\bibitem[\protect\citeauthoryear{Dainotti et al.}{2013a}]{dainotti2013} Dainotti, M.~G., Petrosian, V., Singal, J., Ostrowski, M., 2013a, ApJ, 774, 157 

\bibitem[\protect\citeauthoryear{Dainotti et al.}{2013b}]{dainotti2013b} Dainotti, M.~G., Cardone, V.~F., Piedipalumbo, E., Capozziello, S., 2013b, MNRAS, 2337 

\bibitem[Dainotti et al.(2015a)]{2015ApJ...800...31D} Dainotti, M.~G., Del Vecchio, R., Shigehiro, N., \& Capozziello, S.\ 2015, \apj, 800, 31

\bibitem[\protect\citeauthoryear{Dainotti et al.}{2015b}]{dainotti2015b} Dainotti, M.~G., Petrosian, V., Willingale, R., Obrien, P, Ostrowski, M. \& Nagataki, S, 2015b, MNRAS, 451, 3898 

\bibitem[\protect\citeauthoryear{Dall'Osso et al.}{2011}]{dallosso11} Dall'Osso, S., et al., 2011, A\&A, 526, A121 

\bibitem[\protect\citeauthoryear{Dall'Osso et al.}{2015}]{dallosso15} Dall'Osso, S., et al., 2015, \apj, 798, 25

\bibitem[\protect\citeauthoryear{Duffell \& MacFadyen}{2015}]{dmf15} {Duffell}, P.~C. \& {MacFadyen}, A.~I., 2015, ApJ 806, 205

\bibitem[\protect\citeauthoryear{Duncan \& Thompson}{1992}]{duncan1992} Duncan R.~C. \&Thompson C., 1992, ApJ, 392, L9 

\bibitem[\protect\citeauthoryear{Evans et al.}{2009}]{evans2009} Evans P.~A., et al., 2009, MNRAS, 397, 1177

\bibitem[Efron \& Petrosian(1992)]{1992ApJ...399..345E} Efron, B., \& Petrosian, V.\ 1992, \apj, 399, 345 

\bibitem[Fong et al.(2012)]{2012ApJ...756..189F} Fong, W., et al.\ 2012, \apj, 756, 189 

\bibitem[Giacomazzo \& Perna(2013)]{giacomazzo13} Giacomazzo, B., \& Perna, R.\ 2013, \apjl, 771, 26 

\bibitem[\protect\citeauthoryear{{Goussard}, J.-O. } {et~al.}{2001}]{Gou98} {{Goussard}, J.-O. and {Haensel}, P. \&  {Zdunik}, J.~L.} 1998, A\&A 330, 1005

\bibitem[Graham \& Fruchter (2013)]{graham13} Graham, J. F. \& Fruchter, A. S. 2013, \apj, 774, 119 

\bibitem[Guetta et al.(2004)]{guetta04} Guetta, D., Perna, R., Stella, L., Vietri, M. \ 2004, \apjl, 615, 73 

\bibitem[Guetta \& Della Valle(2007)]{2007ApJ...657L..73G} Guetta, D., \& Della Valle, M.\ 2007, \apjl, 657, L73 

\bibitem[Guetta et al.(2005)]{2005ApJ...619..412G} Guetta, D., Piran, T., \& Waxman, E.\ 2005, \apj, 619, 412 

\bibitem[Gull{\'o}n et al.(2015)]{2015arXiv150705452G} Gull{\'o}n, M., Pons, et al.\ 2015, arXiv:1507.05452 

\bibitem[Gull{\'o}n et al.(2014)]{2014MNRAS.443.1891G} Gull{\'o}n, M., Miralles, J.~A., Vigan{\`o}, D., Pons, J.~A.\ 2014, \mnras, 443, 1891 

\bibitem[Jimenez \& Piran(2013)]{2013ApJ...773..126J} Jimenez, R., \& Piran, T.\ 2013, \apj, 773, 126 

\bibitem[Kiziltan et al (2013)]{kiziltan13} Kiziltan, B., Kottas, A., De Yoreo, M., Thorsett, S. E. \ 2013 \apj, 778, 66


\bibitem[Kuijken \& Gilmore(1989)]{1989MNRAS.239..651K} Kuijken, K., \& Gilmore, G.\ 1989, \mnras, 239, 651 

\bibitem[Kumar et al.(2008)]{2008MNRAS.388.1729K} Kumar, P., Narayan, R., 
\& Johnson, J.~L.\ 2008, \mnras, 388, 1729 


\bibitem[Levesque(2014)]{2014PASP..126....1L} Levesque, E.~M.\ 2014, \pasp, 126, 1 

\bibitem[Lien et al.(2014)]{2014ApJ...783...24L} Lien, A., et al.\ 2014, \apj, 783, 24 


\bibitem[Liang et al.(2007)]{liang07} Liang, E., Zhang, B., Virgili, F., Dai, Z. G. 2007, \apj, 662, 1111

\bibitem[L\"u \& Zhang (2014)]{lu2014} L\"u, H-J. \& Zhang, B. 2014, \apj, 785, 74

\bibitem[\protect\citeauthoryear{Lyons et al.}{2010}]{lyons2009} Lyons N., et al., 2010, MNRAS, 402, 705 

\bibitem[{{Margutti et al.}(2013)}]{margutti2013} {Margutti}, R. et al., 2013, MNRAS, 428, 729

\bibitem[{{Mereghetti}(2008)}]{mereghetti08} {Mereghetti}, S. 2008, \aapr, 15, 225

\bibitem[{{Merloni et al.}(2012)}]{erosita} {Merloni}, A., et al. 2012, MPE document (arXiv:1209.3114)


\bibitem[{{{Meszaros}, P. and {Rees}, M.~J.}(1997)}]{mr97} {M{\'e}sz{\'a}ros}, P. \& {Rees}, M.~J., 1997, ApJ 476, 232

\bibitem[\protect\citeauthoryear{Metzger et al.}{2011}]{metzger2011} Metzger B.~D., et al., 2011, MNRAS, 413, 2031 

\bibitem[Modjaz et al. (2008)]{modjaz08} Modjaz, M., et al.,  2008 \ AJ, 135, 1136


\bibitem[Nousek et al.(2006)]{2006ApJ...642..389N} Nousek, J.~A., 
Kouveliotou, C., Grupe, D., et al.\ 2006, \apj, 642, 389 


\bibitem[Panter et al.(2007)]{2007MNRAS.378.1550P} Panter, B., Jimenez, R., Heavens, A.~F., Charlot, S.\ 2007, \mnras, 378, 1550 

\bibitem[Perna \& Loeb (1998)]{perna98} Perna, R. \& Loeb, A. \ 1998, \apj, 509, 85

\bibitem[Popov (2015)]{popov15} Popov, S. B. 2015, proceedings of STARS2015 (arXiv:1507.06127)

\bibitem[{{Pons} {et~al.}(2013){Pons}, {Vigan\`o}, \& {Rea}}]{pvr13} {Pons}, J.~A., {Vigan\`o}, D., \& {Rea}, N. 2013, Nat. Phys., 9, 431


\bibitem[O'Brien et al.(2006)]{2006ApJ...647.1213O} O'Brien, P.~T., 
Willingale, R., Osborne, J., et al.\ 2006, \apj, 647, 1213 


\bibitem[{{Rea} {et~al.}(2008){Rea}, {Zane}, {Turolla}, {Lyutikov},  {et~al.}}]{rea08} {Rea}, N., {et~al.} 2008, \apj, 686, 1245

\bibitem[{{Rea} \& {Esposito}(2011)}]{rea11} {Rea}, N. \& {Esposito}, P. 2011, in High-Energy Emission from Pulsars and their Systems, ed. D.~F. {Torres} \& N.~{Rea}, 247

\bibitem[Robertson \& Ellis (2012)]{robertson12} Robertson, B. E. \&  Ellis, R. S. \ 2012 \apj, 744, 95 

\bibitem[\protect\citeauthoryear{Rowlinson et al.}{2010}]{rowlinson2010} Rowlinson A., et al., 2010, MNRAS, 408, 383

\bibitem[\protect\citeauthoryear{Rowlinson et al.}{2013}]{rowlinson2013} Rowlinson A., et al., 2013, MNRAS, 430, 1061 

\bibitem[\protect\citeauthoryear{Rowlinson et al.}{2014}]{rowlinson2014} Rowlinson A., et al., 2014, MNRAS, 443, 1779

\bibitem[Salvaterra et al (2012)]{salvaterra12} Salvaterra, R. et al. 2012, \apj, 749, 68

\bibitem[\protect\citeauthoryear{Sari, Piran, \& Narayan}{1998}]{sari1998} Sari R., Piran T. \& Narayan R., 1998, ApJ, 497, L17 

\bibitem[Savaglio et al (2012)]{savaglio12} Savaglio, S. et al. 2012, MNRAS 420, 627

\bibitem[Shen et al. (1998)]{shen98} Shen, H., Toki, H., Oyamatsu, K., Sumiyoshi, K. \ 1998, NuPhA, 637, 435

\bibitem[Soderberg et al.(2010)]{2010Natur.463..513S} Soderberg, A.~M., et al.\ 2010, \nat, 463, 513 

\bibitem[Soderberg et al.(2006)]{2006ApJ...638..930S} Soderberg, A.~M., Nakar, E., Berger, E., Kulkarni, S.~R.\ 2006, \apj, 638, 930 

\bibitem[Trenti et al. (2013)]{trenti13} Trenti, M., Perna, R.,  \& Tacchella, S. \ 2013, \apj, 773L, 22

\bibitem[Trenti et al. (2015)]{trenti15} Trenti, M., Perna, R., \& Jimenez, R. \ 2015, \apj, 802, 103

\bibitem[\protect\citeauthoryear{Troja et al.}{2007}]{troja2007} Troja E., et al., 2007, ApJ, 665, 599 

\bibitem[\protect\citeauthoryear{Usov}{1992}]{usov92} Usov V.~V., 1992, Nature, 357, 472 

\bibitem[{{Vigan{\`o}} {et~al.}(2013){Vigan{\`o}}, {Rea}, {Pons},  {et~al.}}]{vigano13} {Vigan{\`o}}, D., {et~al.}, 2013, \mnras, 434, 123

\bibitem[Wanderman \& Piran(2015)]{2015MNRAS.448.3026W} Wanderman, D., \& Piran, T.\ 2015, \mnras, 448, 3026 

\bibitem[Wanderman \& Piran(2010)]{2010MNRAS.406.1944W} Wanderman, D., \& Piran, T.\ 2010, \mnras, 406, 1944 

\bibitem[\protect\citeauthoryear{Willingale et al.}{2007}]{willingale2007} Willingale R., et al., 2007, ApJ, 662, 1093 

\bibitem[\protect\citeauthoryear{Xiao \& Schaefer}{2009}]{X09} {Xiao}, L. \& {Schaefer}, B.~E., 2009, ApJ 707, 387

\bibitem[Yusifov {\& Kuuml}{\c c}{\"u}k(2004)]{2004A&A...422..545Y} Yusifov, I., {\& Kuuml}{\c c}{\"u}k, I.\ 2004, \aap, 422, 545 

\bibitem[\protect\citeauthoryear{Zhang \& M{\'e}sz{\'a}ros}{2001}]{zhang2001} Zhang B., \& M{\'e}sz{\'a}ros P., 2001, ApJ, 552, L35 


\bibitem[Zhang et al.(2006)]{2006ApJ...642..354Z} Zhang, B., Fan, Y.~Z., 
Dyks, J., et al.\ 2006, \apj, 642, 354 


\bibitem[Zane et al.(2009)]{2009MNRAS.398.1403Z} Zane, S., Rea, N., Turolla, R., Nobili, L.\ 2009, \mnras, 398, 1403 


\end{thebibliography}
\end{document}